\documentclass[11pt]{article}

\usepackage{graphicx}
\usepackage{psfrag}
\usepackage[small]{caption2}
\usepackage[below]{placeins}
\usepackage{flafter}
\usepackage{amssymb}
\usepackage{amsmath}

\newcommand{\non}{\nonumber}
\def\be{\begin{equation}}
\def\ee{\end{equation}}
\def\bea{\begin{eqnarray}}
\def\eea{\end{eqnarray}}

\def\sbra#1{\left\langle #1\right|}           
\def\sket#1{\left| #1\right\rangle}           
\def\sVEV#1{\left\langle #1\right\rangle}     
\def\ha{\frac12}                              

\def\a{\alpha}
\def\b{\beta}

\def\p{\pi}           
\def\q{\theta}        
\def\s{\sigma}        

\begin{document}

\title{{\bf Grover's Algorithm: Quantum Database Search\footnote{Contents based
on lecture notes from graduate courses in Quantum Computation given at LNCC.}}}

\author{C. Lavor \\
{\small Instituto de Matem\'{a}tica e Estat\'{\i}stica }\\ {\small
Universidade do Estado do Rio de Janeiro - UERJ}\\ {\small Rua
S\~{a}o Francisco Xavier, 524, 6${{}^o}$%
andar, bl. D, sala 6018,}\\ {\small Rio de Janeiro, RJ, 20550-900,
Brazil}\\ \textit{e-mail: carlile@ime.uerj.br} \and
\\
L.R.U. Manssur, R. Portugal \\
{\small Coordena\c{c}\~{a}o de Ci\^encia da Computa\c{c}\~{a}o}\\
{\small Laborat\'{o}rio Nacional de Computa\c{c}\~{a}o
Cient\'{\i}fica - LNCC}\\ {\small Av. Get\'{u}lio Vargas 333,
Petr\'{o}polis, RJ, 25651-070, Brazil}\\ \textit{e-mail:
\{leon,portugal\}@lncc.br} } \maketitle

\begin{abstract}

{\normalsize \noindent We review Grover's algorithm by means of a
detailed geometrical interpretation and a worked out example. Some
basic concepts of Quantum Mechanics and quantum circuits are also
reviewed. This work is intended for non-specialists which have
basic knowledge on undergraduate Linear Algebra. }

\end{abstract}

\section{Introduction}

The development of quantum software and hardware is an exciting
new area posing extremely difficult challenges for researchers all
over the world. It promises a new era in Computer Science, but it
is not clear at all whether it will be possible to build a
hardware of reasonable size. Quantum hardware of macroscopic sizes
suffer the decoherence effect which is an unsurmountable tendency
to behave classically.

An important landmark in hardware development is the experience
performed at IBM's lab in San Jose, California, which factored the
number 15 into its prime factors using a quantum algorithm (Shor's
algorithm \cite{shor}) executed in a molecule, perfluorobutadienyl
iron complex \cite{exp}. This ``quantum computer'' has seven
``quantum bits''. Such insignificant amount of bits that could be
manipulated in the experience shows the challenge in hardware
development.

Quantum software development is facing difficulties too, though
the superiority of quantum circuits over classical ones was
already established. In this context, Grover's algorithm
\cite{grover1,grover2} plays an important role, since it provides
a proof that quantum computers are faster than classical ones in
database searching. The best classical algorithm for an
unstructured database search has complexity $O(N)$, without
possibility of further improvement, while the best quantum
algorithm has complexity $O(\sqrt N )$.

Historically, Deutsch's algorithm \cite{deutsch1} was the first
example of a quantum circuit faster than its classical
counterpart, while Bernstein and Vazirani \cite{vazirani} and
Simon \cite{simon} provided the first examples of quantum
algorithms exponentially faster than their classical counterparts.
These algorithms determine some sort of functions' periods and
their only application seems to be for proving that quantum
circuits are faster than classical ones.

Some of the most impressive results so far in quantum computation
are the Shor's algorithms \cite{shor} for factoring integers and
for finding discrete logarithms, which also provided an
exponential speed up over the known classical algorithms. Shor's
results attracted a lot of attention because they render most of
current cryptography methods useless, if we assume it is possible
to build quantum hardware of reasonable size.

This work is an introductory review of Grover's algorithm. We have
put all our efforts to write as clear as possible for
non-specialists. We assume familiarity with undergraduate Linear
Algebra, which is the main mathematical basis of Quantum
Mechanics. Some previous knowledge of Quantum Mechanics is
welcome, but not necessary for a persistent reader.
The reader can find further material in
\cite{aharonov,chuang,preskill,bennett,boyer,brassard}.

Section 2 reviews basic notions about classical computers preparing for the quantum
generalization, which is presented in Section 3. Section 4 introduces the
notion of quantum circuits and presents some basic examples. Section
5 describes Grover's algorithm. Sections 6 and 8 give details of the geometrical
interpretation while Section 7 presents a worked out example. Finally, Section 9
shows the decomposition of Grover's circuit in terms of the universal gates.

\section{The Classical Computer}

A classical computer can be understood in a very broad
sense as a machine that reads a certain amount of data, encoded as
zeroes and ones, performs calculations, and prints in the end output
data as zeroes and ones again. Zeroes and ones are states of some physical
quantity, the electric potential in classical computers. Internally, a zero
is a state of low electric potential and a one is a state of high electric
potential. This point is crucial to the generalization we will discuss
ahead.

\begin{figure}
    \setcaptionmargin{.5in}
    \centering
    \psfrag{bit0}{bit$_0$}
    \psfrag{bit1}{bit$_1$}
    \psfrag{bitn-1}{bit$_{n-1}$}
    \psfrag{f}{$f$}
    \includegraphics[]{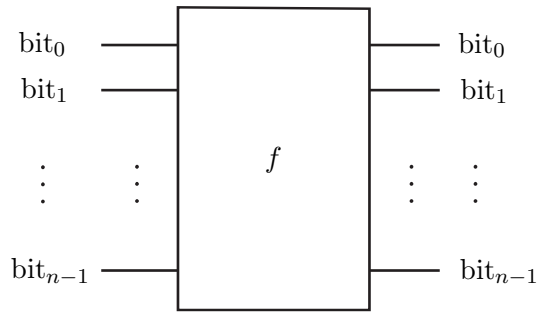}
    \caption[text for list of figures]{Outline of the classical computer.}
    \label{fig01}
\end{figure}
Zeroes and ones form a binary number which can be converted to
decimal notation. Let us think of the computer as calculating a
function
\begin{equation*}
f:\{0,...,N-1\}\rightarrow \{0,...,N-1\},
\end{equation*}
where $N$ is a number of the form $2^{n}$ ($n$ is the number of
bits in the computer memory). We assume without loss of generality
that the domain and codomain of $f$ are of the same size ($f$ is a
function because one input cannot generate two or more outputs).
We represent the calculation process in Fig. \ref{fig01}, where on
the left hand side we have the value of each bit (zero or one).
The process of calculation goes from left to right, and the output
is stored in the same bits on the right hand side.

\begin{figure}
    \setcaptionmargin{.5in}
    \centering
    \includegraphics[width=5in]{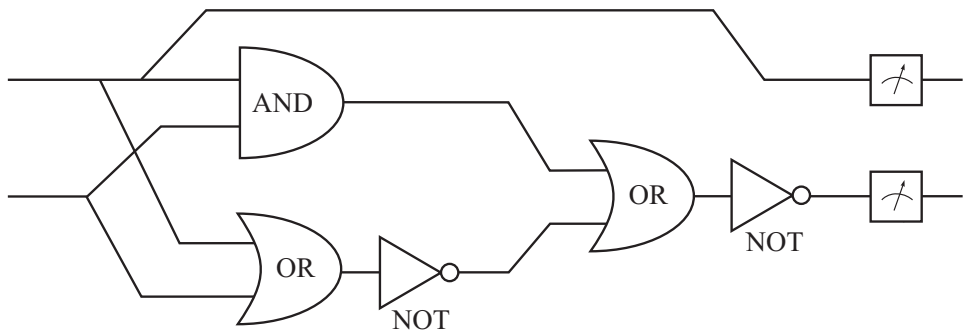}
    \caption[text for list of figures]{The circuit to add two one-bit
numbers modulo 2.}
    \label{fig02}
\end{figure}
Usually $f$ is given in terms of elementary blocks that can be
implemented in practice using transistors and other electrical
devices. The blocks are the AND, OR and NOT gates, known as
universal gates (This set could be reduced further since OR can be
written in terms of AND and NOT). For example, the circuit to add
two one-bit numbers modulo 2 is given in Fig. \ref{fig02}. The
possible inputs are $00$, $01$, $10$, $11$, and the corresponding
outputs are $00$, $01$, $11$, $10$. The inputs are prepared
creating electric potential gaps, which create electric currents
that propagate through the wires towards right. The gates are
activated as time goes by. The meter symbols on the right indicate
that measurements of the electric potential are performed, which
tell whether the output value of each bit is zero or one. The
second bit gives the result of the calculation. The wire for the
output of the first bit is artificial and unnecessary; at this
point, it is there simply to have the codomain size of the
function $f$ equal to the domain size. This circuit, without the
first bit output wire, is the circuit for the XOR (exclusive OR)
gate in terms of the universal gates.

The circuit of Fig. \ref{fig02} is irreversible, since the gates
AND and OR are irreversible. If the output of the AND gate is 0,
nobody can tell what was the input, and similarly when the output
of the OR gate is 1. This means that the physical theory which
describes the processes in Fig. \ref{fig02} must be irreversible.
Then, the AND and OR gates cannot be straightforwardly generalized
to quantum gates, which must be reversible ones.

\begin{figure}
    \setcaptionmargin{.5in}
    \centering
    \psfrag{a}{$a$}
    \psfrag{b}{$b$}
    \psfrag{a+b(mod2)}{ $a+b \mbox{ (mod 2)}$}
    \includegraphics[]{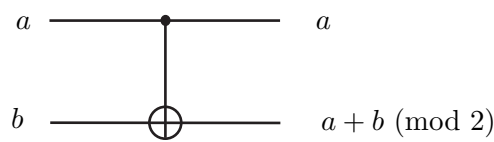}
    \caption[text for list of figures]{Classical controlled-NOT (CNOT) gate.}
    \label{fig03}
\end{figure}

However, the circuit of Fig. \ref{fig02} can be made reversible.
Although the following description is unnecessary from the
classical point of view, it helps the quantum generalization
performed in the next sections. We employ the controlled-NOT
(CNOT) gate of Fig. \ref{fig03}.
The bits $a$ and $b$ assume values either 0 or 1. The value of the
first bit (called the control bit) never changes in this gate; the
second bit (called the target bit) is flipped only if $a=1$. If
$a=0$, nothing happens to both bits. The gate $\oplus$ is a NOT
gate controlled by the value of the first bit. Now it is easy to
verify that the value of the second bit for this gate is
$a+b\mbox{ (mod 2)}$. The CNOT gate is not a universal building
block for classical circuits, but its quantum counterpart is a
basic block of quantum circuits.

\begin{figure}
    \setcaptionmargin{.5in}
    \centering
    \psfrag{a}{$a$}
    \psfrag{b}{$b$}
    \psfrag{c}{$c$}
    \psfrag{c+ab(mod2)}{$c+ab \mbox{ (mod 2)}$}
    \includegraphics[]{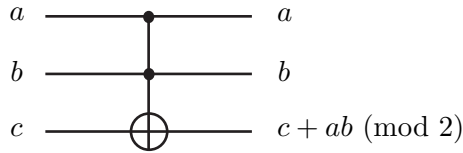}
    \caption[text for list of figures]{Classical Toffoli gate.}
    \label{fig04}
\end{figure}

We have described the reversible counterpart of the XOR gate. What
is the reversible counterpart of the AND gate? The answer employs
the {\em Toffoli} gate (Fig. \ref{fig04}) which is a
generalization of the CNOT gate with two control bits instead of
one.
The value of the third bit (target) is inverted only if both $a$
and $b$ are 1, otherwise it does not change. The following table
describes all possible inputs and the corresponding outputs:
\begin{eqnarray*}
000 &\rightarrow &000 \\
001 &\rightarrow &001 \\
010 &\rightarrow &010 \\
011 &\rightarrow &011 \\
100 &\rightarrow &100 \\
101 &\rightarrow &101 \\
110 &\rightarrow &111 \\
111 &\rightarrow &110
\end{eqnarray*}
The AND gate can be replaced by the Toffoli gate simply by taking
$c=0$. The output of the third bit is then $a$ AND $b$ (The
reversible circuit for the OR gate is a little cumbersome because
it requires more than one Toffoli gate, so we will not describe it
here).

Another feature implicit in Fig. \ref{fig02} that cannot be
performed in quantum circuits is FANOUT. Note that there are
bifurcations of wires; there is no problem to do this classically.
However, this is forbidden in quantum circuits, due to the ``no
cloning'' theorem (see \cite{preskill} p.162). Classical FANOUT
can be obtained from the CNOT gate by taking $b=0$. The value of
the first bit is then duplicated.

Consider again Fig. \ref{fig01}. If the computer has $n$ bits,
there are $2^{n}$ possible inputs. For each input there are $2^n$
possible outputs, therefore the number of possible functions $f$
that can be calculated is $2^{n 2^{n}}$. All these functions can
be reduced to circuits using the universal gates. That is what a
classical computer can do: calculate $2^{n 2^{n}}$ functions. This
number is astronomical for computers with $1$ gigabyte, that is a
typical memory size for good personal computers nowadays.

Another important issue is how fast can the computer calculate
these functions. The answer depends on the number of gates used in
the circuit for $f$. If the number of elementary gates increases
polynomially with $n$, we say that the circuit is ``efficient''.
If the number of gates increases exponentially with $n$, the
circuit is ``inefficient''. This is a very coarse method to
measure the efficiency, but it is useful for theoretical analysis
when $n$ is large. Note that we are thinking of computers of
variable size, which is not the case in practice. In fact, instead
of referring to actual computers, it is better to use a Turing
machine, which is an abstract model for computers and softwares as
a whole \cite{papadimitriou}. Similarly, quantum computers and
their softwares are abstractly modeled as the quantum Turing
machine \cite{deutsch2, vazirani}. The classical theory of
complexity classes and its quantum counterpart address this kind
of problems.

All calculations that can be performed in a classical computer can
also be performed in a quantum computer. One simply replaces the
irreversible gates of the classical computer with their reversible
counterparts. The new circuit can be implemented in a quantum
computer. But there is no advantage in this procedure: why build a
very expensive quantum machine which behaves classically? The
appeal of quantum computation is the possibility of quantum
algorithms faster than classical ones. The quantum algorithms must
use quantum features not available in classical computers, such as
quantum parallelism and entanglement, in order to enhance the
computation. On the other hand, a na\"ive use of quantum features
does not guarantee any improvements. So far, there are only two
classes of successful quantum algorithms: the database search
algorithms and the algorithms for finding the generators of a
normal subgroup of a given group. Shor's algorithms for integer
factorization and discrete logarithm are special cases of this
latter class.

\section{The Quantum Computer}

\begin{figure}
    \setcaptionmargin{.5in}
    \centering
    \psfrag{1}{$\sket 1$}
    \psfrag{psi}{$\sket \psi$}
    \psfrag{0}{$\sket 0$}
    \includegraphics[height=6cm]{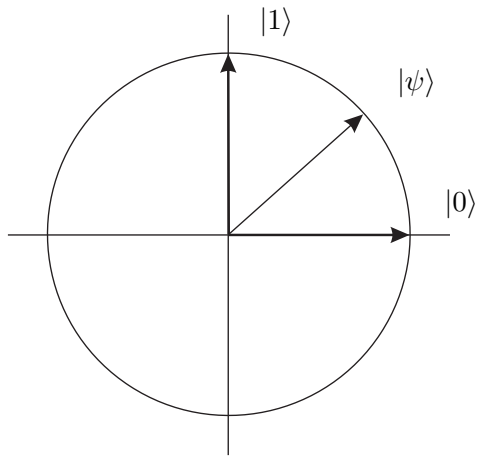}
    \caption[text for list of figures]{Computational basis for the case
$\a$, $\b$ real. In the general case ($\a$, $\b$ complex) there is still a
geometrical representation called the Bloch sphere \cite{chuang}.}
    \label{fig05}
\end{figure}

In quantum computers, one is allowed to use quantum states instead
of classical ones. So, the electric potential can be replaced by
some quantum state: the \textit{quantum bit} ({\em qubit} for
short). Just as a bit has a state $0$ or $1$, a qubit also has a
state $|0\rangle $ or $|1\rangle . $ This is called the {Dirac
notation} and it is the standard notation for states in Quantum
Mechanics. The difference between bits and qubits is that a qubit
$|\psi \rangle $ can also be in a linear combination of states
$|0\rangle $ and $|1\rangle $:
\begin{equation}
|\psi \rangle =\alpha |0\rangle +\beta |1\rangle .  \label{a}
\end{equation}
This state is called a {\em superposition} of the states
$|0\rangle $ and
$|1\rangle $ with {\em amplitudes} $\alpha $ and $\beta $ ($\alpha $ and $%
\beta $ are complex numbers). Thus, the state $|\psi \rangle $ is
a vector in a two-dimensional complex vector space, where the
states $|0\rangle $ and $|1\rangle $ form an orthonormal basis,
called the {computational basis} (see Fig. \ref{fig05} in the real
case).

The state $|0\rangle $ is not the zero vector, but simply the first vector
of the basis. 
The matrix representations of the vectors $%
|0\rangle $ and $|1\rangle $ are given by
\begin{equation*}
|0\rangle =\left[
\begin{array}{c}
1 \\
0
\end{array}
\right] \;\;\;\mbox{ and }\;\;\;
|1\rangle =\left[
\begin{array}{c}
0 \\
1
\end{array}
\right] .
\end{equation*}
What is the interpretation of $\alpha $ and $\beta $ in Eq.
(\ref{a})? Quantum mechanics tells us that if one measures the
state $|\psi \rangle $
one gets either $|0\rangle ,$ with probability $|\alpha |^{2},$ or $%
|1\rangle ,$ with probability $|\beta |^{2}$. That is, measurement
changes the state of a qubit. In fact, any attempt to find out the
amplitudes of the state $|\psi \rangle $ produces a nondeterministic
collapse of the superposition to either $|0\rangle $ or $|1\rangle $. If $%
|\alpha |^{2}$ and $|\beta |^{2}$ are probabilities and there are only two
possible outputs, then
\begin{equation}
|\alpha |^{2}+|\beta |^{2}=1.  \label{b}
\end{equation}
Calculating the norm of $|\psi \rangle $, Eq. (\ref{b}) gives
\begin{equation*}
||\;|\psi \rangle \;||=\sqrt{|\alpha |^{2}+|\beta |^{2}} = 1.
\end{equation*}

If a qubit is in state $|\psi \rangle $ given by Eq. (\ref{a}),
there are two ways it can interact. The first one is a
measurement. This forces the state $|\psi \rangle $ to collapse to
either $|0\rangle $ or $|1\rangle $ with probabilities $|\alpha |^{2}$ and $%
|\beta |^{2}$, respectively. Note that the measurement does not
give the value of $\alpha $ and $\beta $. They are inaccessible
via measurements unless one has many copies of the same state. The
second kind of interaction does not give any information about the
state. In this case, the values of $\alpha $ and $\beta $ change
keeping the constraint (\ref{b}). The most general transformation
of this kind is a linear transformation $U$ that takes unit
vectors into unit vectors. Such transformation is called
\textit{unitary} and can be defined by
\begin{equation*}
U^{\dagger }U=UU^{\dagger }=I,
\end{equation*}
where $U^{\dagger }$ $=(U^{\ast })^{T}$ ($\ast $ indicates complex
conjugation and $T$ indicates the transpose operation).

To consider multiple qubits it is necessary to introduce the concept of
\textit{tensor product}. Suppose $V$ and $W$ are complex vector spaces of
dimensions $m$ and $n$, respectively. The tensor product $V\otimes W$ is an $%
mn$-dimensional vector space. The elements of $V\otimes W$ are linear
combinations of $\ $tensor products $|v\rangle \otimes |w\rangle $,
satisfying the following properties ($z\in \mathbb{C}$, $|v\rangle
,|v_{1}\rangle ,|v_{2}\rangle \in V$, and $|w\rangle ,|w_{1}\rangle
,|w_{2}\rangle \in W$):

\begin{enumerate}
\item  $z(|v\rangle \otimes |w\rangle )=(z|v\rangle )\otimes |w\rangle
=|v\rangle \otimes (z|w\rangle ),$

\item  $(|v_{1}\rangle +|v_{2}\rangle )\otimes |w\rangle =(|v_{1}\rangle
\otimes |w\rangle )+(|v_{2}\rangle \otimes |w\rangle ),$

\item  $|v\rangle \otimes (|w_{1}\rangle +|w_{2}\rangle )=(|v\rangle \otimes
|w_{1}\rangle )+(|v\rangle \otimes |w_{2}\rangle ).$
\end{enumerate}

\noindent We use also the notations $|v\rangle |w\rangle ,$ $|v,w\rangle $ or $%
|vw\rangle $ for the tensor product $|v\rangle \otimes |w\rangle
$. Note that the tensor product is non-commutative, so the
notation must preserve the ordering.

Given two linear operators $A$\ and $B$\ defined on the vector spaces $V$\
and $W$, respectively, we can define the linear operator $A\otimes B$\ on $%
V\otimes W$\ as
\begin{equation}
(A\otimes B)(|v\rangle \otimes |w\rangle )=A|v\rangle \otimes
B|w\rangle,\label{c}
\end{equation}
where $|v\rangle \in V$ and $|w\rangle \in W$. The matrix
representation of $A\otimes B$ is given by
\begin{equation}
A\otimes B=\left[
\begin{array}{rrr}
A_{11}B & \cdot \cdot \cdot & A_{1m}B \\
\vdots & \ddots & \vdots \\
A_{m1}B & \cdot \cdot \cdot & A_{mm}B
\end{array}
\right] ,  \label{c1}
\end{equation}
where $A$ is an $m\times m$ matrix and $B$ is a $n\times n$ matrix
(We are using the same notation for the operator and its matrix
representation) . So the matrix $A\otimes B$ has dimension
$mn\times mn$. For example, given
\begin{equation*}
A=\left[
\begin{array}{rr}
0 & 1 \\
1 & 0
\end{array}
\right] \;\;\;\text{  and  }\;\;\;B=\left[
\begin{array}{rrr}
1 & 0 & 0 \\
0 & 1 & 0 \\
0 & 0 & 1
\end{array}
\right] ,
\end{equation*}
the tensor product $A\otimes B$ is
\begin{equation*}
A\otimes B=\left[
\begin{array}{rr}
0 & 1 \\
1 & 0
\end{array}
\right] \otimes \left[
\begin{array}{rrr}
1 & 0 & 0 \\
0 & 1 & 0 \\
0 & 0 & 1
\end{array}
\right] =\left[
\begin{array}{rrrrrr}
0 & 0 & 0 & 1 & 0 & 0 \\
0 & 0 & 0 & 0 & 1 & 0 \\
0 & 0 & 0 & 0 & 0 & 1 \\
1 & 0 & 0 & 0 & 0 & 0 \\
0 & 1 & 0 & 0 & 0 & 0 \\
0 & 0 & 1 & 0 & 0 & 0
\end{array}
\right] .
\end{equation*}
The formula (\ref{c1}) can also be used for non square matrices,
such as the tensor product of two vectors. For example, the tensor
product $|0\rangle \otimes |1\rangle $ is given by
\begin{equation*}
|0\rangle \otimes |1\rangle =|01\rangle =\left[
\begin{array}{r}
1 \\
0
\end{array}
\right] \otimes \left[
\begin{array}{r}
0 \\
1
\end{array}
\right] =\left[
\begin{array}{r}
0 \\
1 \\
0 \\
0
\end{array}
\right] .
\end{equation*}
The notations $|\psi \rangle ^{\otimes k}$ and $A^{\otimes k}$
mean $|\psi \rangle $ and $A$ tensored with themselves $k$ times,
respectively.

The general state $|\psi \rangle $ of two qubits is a superposition of the
states $|00\rangle ,$ $|01\rangle $, $|10\rangle $, and $|11\rangle $:
\begin{equation}
|\psi \rangle =\alpha |00\rangle +\beta |01\rangle +\gamma |10\rangle
+\delta |11\rangle ,  \label{d}
\end{equation}
with the constraint
\begin{equation*}
|\alpha |^{2}+|\beta |^{2}+|\gamma |^{2}+|\delta |^{2}=1.
\end{equation*}
Regarding the zeroes and ones as constituting the binary expansion of an
integer, we can replace the representations of states
\begin{equation*}
|00\rangle ,\text{ }|01\rangle ,\text{ }|10\rangle ,\text{ }|11\rangle ,
\end{equation*}
by the shorter forms
\begin{equation*}
|0\rangle ,\text{ }|1\rangle ,\text{ }|2\rangle ,\text{
}|3\rangle,
\end{equation*}
in decimal notation.

In general, the state $|\psi \rangle $ of $n$ qubits is a
superposition of the $2^{n}$ states $|0\rangle ,$ $|1\rangle ,$ $...,$ $%
|2^{n}-1\rangle $:
\begin{equation*}
|\psi \rangle =\underset{i=0}{\overset{2^{n}-1}{\sum }}\alpha
_{i}|i\rangle ,
\end{equation*}
with amplitudes $\alpha _{i}$ constrained to
\begin{equation*}
\underset{i=0}{\overset{2^{n}-1}{\sum }}|\alpha _{i}|^{2}=1.
\end{equation*}
The orthonormal basis $\{ \sket{0}, \dots, \sket{2^n-1} \}$ is
called {\it computational basis}. As before, a measurement of a
generic state $\sket \psi$ yields the result $\sket{i_0}$ with
probability $|\alpha _{i_0}|^{2}$, where $0 \leq i_0 < N$.
Usually, the measurement is performed qubit by qubit yielding
zeroes or ones that are read together to form $i_0$. We stress
again a very important feature of the measurement process. The
state $\sket \psi$ as it is before measurement is inaccessible
unless it is in the computational basis. The measurement process
inevitably disturbs $\sket \psi$ forcing it to collapse to one
vector of the computational basis. This collapse is
non-deterministic, with the probabilities given by the squared
norms of the corresponding amplitudes in $\sket \psi$.

If we have two qubits, one in the state
\begin{equation*}
|\varphi \rangle =a|0\rangle +b|1\rangle
\end{equation*}
and the other in the state
\begin{equation*}
|\psi \rangle =c|0\rangle +d|1\rangle ,
\end{equation*}
then the state of the pair $|\varphi \rangle |\psi \rangle $ is the tensor
product
\begin{eqnarray}
|\varphi \rangle \otimes |\psi \rangle &=&(a|0\rangle +b|1\rangle )\otimes
(c|0\rangle +d|1\rangle )  \label{e} \\
&=&ac|00\rangle +ad|01\rangle +\ bc|10\rangle +bd|11\rangle .  \notag
\end{eqnarray}
Note that a general $2$-qubit state (\ref{d}) is of the form (\ref{e}) if
and only if
\begin{eqnarray*}
\alpha &=&ac, \\
\beta &=&ad, \\
\gamma &=&bc, \\
\delta &=&bd.
\end{eqnarray*}
>From these equalities we have that a general $2$-qubit state
(\ref{d}) is of the form (\ref{e}) if and only if
\begin{equation*}
\alpha \delta =\beta \gamma .
\end{equation*}
Thus, the general $2$-qubit state is
not a product of two $1$-qubit states. Such non-product states of two or more
qubits are called \textit{entangled} states, for example, $(\sket{00} +
\sket{11}) / \sqrt 2$.

There is an \textit{inner product} between two $n$-qubit states
$|\varphi \rangle $ and $|\psi \rangle $, written in the form
$\langle \varphi |\psi \rangle $, which is defined by the
following rules in a complex vector space $V$:

\begin{enumerate}
\item  $\langle \psi |\varphi \rangle =\langle \varphi |\psi \rangle ^{\ast
},$

\item  $\langle \varphi |(a|u\rangle +b|v\rangle )\rangle =a\langle \varphi
|u\rangle +b\langle \varphi |v\rangle ,$

\item  $\langle \varphi |\varphi \rangle >0$ if $|\varphi \rangle \neq 0,$
\end{enumerate}

\noindent where $a,b\in \mathbb{C}$ and $|\varphi \rangle ,|\psi
\rangle ,|u\rangle ,|v\rangle \in V.$ The {\em norm} of a vector
$|\varphi \rangle $ is given by
\begin{equation*}
||\;|\varphi \rangle \;||=\sqrt{\langle \varphi |\varphi \rangle }.
\end{equation*}

The notation $\langle \varphi |$ is used for the \textit{%
dual vector} to the vector $|\varphi \rangle $. The dual is a linear
operator from the vector space $V$ to the complex numbers, defined by
\begin{equation*}
\langle \varphi |(|v\rangle )=\langle \varphi |v\rangle , \;\;\; \text{   }
\forall
|v\rangle \in V.
\end{equation*}

Given two vectors $|\varphi \rangle $ and $|\psi \rangle $ in a vector space
$V$, there is also an \textit{outer product} $|\psi \rangle \langle \varphi
| $, defined as a linear operator on $V$ satisfying
\begin{equation*}
(|\psi \rangle \langle \varphi |)|v\rangle =|\psi \rangle \langle \varphi
|v\rangle , \;\;\; \text{   }\forall |v\rangle \in V.
\end{equation*}

If $|\varphi \rangle =a |0\rangle +b |1\rangle $ and $|\psi \rangle
=c |0\rangle +d |1\rangle $, then the matrix representations for
inner and outer products are:
\begin{eqnarray*}
\langle \varphi |\psi \rangle &=&\left[
\begin{array}{rr}
a ^{\ast } & b ^{\ast }
\end{array}
\right] \left[
\begin{array}{r}
c \\
d
\end{array}
\right] =a ^{\ast }c +b ^{\ast }d, \\
|\varphi \rangle \langle \psi | &=&\left[
\begin{array}{r}
a \\
b
\end{array}
\right] \left[
\begin{array}{rr}
c ^{\ast } & d ^{\ast }
\end{array}
\right] =\left[
\begin{array}{rr}
ac ^{\ast } & ad ^{\ast } \\
bc ^{\ast } & bd ^{\ast }
\end{array}
\right] .
\end{eqnarray*}
Notice the complex conjugation in the process of taking the dual.
\begin{figure}
    \setcaptionmargin{.5in}
    \centering
    \psfrag{bit0}{$\sket {\psi_1}$}
    \psfrag{bit1}{$\sket {\psi_2}$}
    \psfrag{bitn-1}{$\sket {\psi_n}$}
    \psfrag{bit0p}{$\sket {\psi^\prime_1}$}
    \psfrag{bit1p}{$\sket {\psi^\prime_2}$}
    \psfrag{bitn-1p}{$\sket {\psi^\prime_n}$}
    \psfrag{psi}[][]{$\sket {\psi}$}
    \psfrag{f}{$U$}
    \includegraphics[]{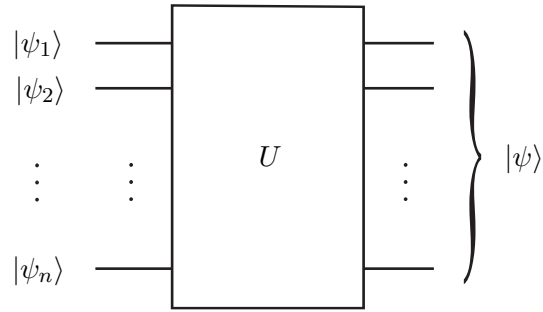}
    \caption[text for list of figures]{The sketch of the quantum computer. We
    consider the input non-entangled, which is reasonable in general. On the other hand,
    the output is entangled in general. The measurement of the state $\sket \psi$,
    not shown here, returns zeroes and ones.}
    \label{fig06}
\end{figure}

After the above review, we are ready to outline the quantum
computer. Fig. \ref{fig06} is the generalization of Fig.
\ref{fig01} to the quantum case. The function $f$ is replaced by a
unitary operator $U$ and classical bits are replaced by quantum
bits, where each one has a state $|\psi _{i}\rangle .$ In Fig.
\ref{fig06}, we are taking a non-entangled input, what is quite
reasonable. In fact, $\sket{\psi_i}$ is either $\sket 0$ or $\sket
1$ generally. $\sket \psi$ on the right hand side of Fig.
\ref{fig06} is the result of the application of $U$ on the input.
The last step is the measurement of the states of each qubit,
which returns zeroes and ones that form the final result of the
quantum calculation. Note that there is, in principle, an infinite
number of possible operators $U$, which are unitary $2^n\times
2^n$ matrix, with continuous entries. In particular, one must take
errors into account, which reduces the number of implementable
circuits. But even in this case, the number of degrees of freedom
is greater than in the classical case.

Similarly to the classical case, the operator $U$ is in general written in
terms of gates forming a quantum circuit, which is the topic of the next
section.

\section{Quantum Circuits}

%
\begin{figure}
    \setcaptionmargin{.5in}
    \centering
    \psfrag{psi}{$\sket \psi$}
    \psfrag{Xpsi}{$X \sket \psi$}
    \includegraphics{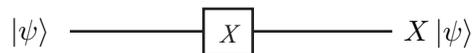}
    \caption[text for list of figures]{Quantum NOT gate.}
    \label{fig07}
\end{figure}
Let us start with one-qubit gates. In the classical case there is
only one possibility, which is the NOT gate, like the ones used in
Fig. \ref{fig02}. The straightforward generalization to the
quantum case is given in Fig. \ref{fig07}, where $X$ is the
unitary operator
\begin{equation*}
X=\left[
\begin{array}{cc}
0 & 1 \\
1 & 0
\end{array}
\right] .
\end{equation*}
So, if $|\psi \rangle $ is $|0\rangle ,$ the output is $|1\rangle $ and
vice-versa. But now we have a situation with no classical counterpart. The
state $|\psi \rangle $ can be a superposition of states $|0\rangle $ and $%
|1\rangle $. The general case is given in Eq. (\ref{a}). The
output in this case is $\alpha |1\rangle +\beta |0\rangle $.

The gate $X$ is not the only one-qubit gate. There are infinitely
many, since there are an infinite number of $2\times 2$ unitary
matrices. In principle, any unitary operation can be implemented
in practice. The {\em Hadamard} gate is another important
one-qubit gate, given by
\begin{equation*}
H=\frac{1}{\sqrt{2}}\left[
\begin{array}{rr}
1 & 1 \\
1 & -1
\end{array}
\right] .
\end{equation*}
It is easy to see that
\begin{eqnarray*}
H|0\rangle  &=&\frac{|0\rangle +|1\rangle }{\sqrt{2}}, \\
H|1\rangle  &=&\frac{|0\rangle -|1\rangle }{\sqrt{2}}.
\end{eqnarray*}
If the input is $|0\rangle $, the Hadamard gate creates a superposition of
states with equal weights. This is a general feature, valid for two or more
qubits. Let us analyze the $2$-qubit case.

The first example of a $2$-qubit gate is $H\otimes H$:
\begin{eqnarray*}
H^{\otimes 2}|0\rangle |0\rangle  &=&(H\otimes H)(|0\rangle \otimes
|0\rangle )=H|0\rangle \otimes H|0\rangle  \\
&=&\left( \frac{|0\rangle +|1\rangle }{\sqrt{2}}\right) \otimes \left( \frac{%
|0\rangle +|1\rangle }{\sqrt{2}}\right)  \\
&=&\frac{1}{2}(|0\rangle |0\rangle +|0\rangle |1\rangle +|1\rangle |0\rangle
+|1\rangle |1\rangle ) \\
&=&\frac{1}{2}(|0\rangle +|1\rangle +|2\rangle +|3\rangle ).
\end{eqnarray*}
The result is a superposition of all basis states with equal
weights. More generally, the Hadamard operator applied to the
$n$-qubit state $|0\rangle$ is
\begin{equation*}
H^{\otimes n}|0\rangle=\frac{1}{\sqrt{2^{n}}}\underset{i=0}{\overset{%
2^{n}-1}{\sum }}|i\rangle.
\end{equation*}
Thus, the tensor product of $n$ Hadamard operators produces an
equally weighted \ superposition of all computational basis
states, when the input is the state $|0\rangle.$

Another important $2$-qubit quantum gate is the CNOT gate, which
is the quantum generalization of the classical gate described
earlier (Fig. \ref{fig03}). It has two input qubits, the control
and the target qubit, respectively. The target qubit is flipped
only if the control qubit is set to 1, that is,
\bea
|00\rangle & \rightarrow & |00\rangle, \non \\
|01\rangle & \rightarrow & |01\rangle, \label{cnot}\\
|10\rangle & \rightarrow & |11\rangle, \non \\
|11\rangle & \rightarrow & |10\rangle . \non
\eea
The action of the CNOT gate can also be represented by
\begin{equation*}
|a,b\rangle \rightarrow |a,a\oplus b\rangle ,
\end{equation*}
where $\oplus $ is addition modulo 2. Now, let us obtain its
matrix representation. We know that
\begin{eqnarray}
|00\rangle  &=&|0\rangle \otimes |0\rangle =\left[
\begin{array}{r}
1 \\
0
\end{array}
\right] \otimes \left[
\begin{array}{r}
1 \\
0
\end{array}
\right] =\left[
\begin{array}{r}
1 \\
0 \\
0 \\
0
\end{array}
\right] ,  \notag
\end{eqnarray}
\begin{eqnarray}
|01\rangle  &=&|0\rangle \otimes |1\rangle =\left[
\begin{array}{r}
1 \\
0
\end{array}
\right] \otimes \left[
\begin{array}{r}
0 \\
1
\end{array}
\right] =\left[
\begin{array}{r}
0 \\
1 \\
0 \\
0
\end{array}
\right] , \label{vectors}
\end{eqnarray}
\begin{eqnarray}
|10\rangle  &=&|1\rangle \otimes |0\rangle =\left[
\begin{array}{r}
0 \\
1
\end{array}
\right] \otimes \left[
\begin{array}{r}
1 \\
0
\end{array}
\right] =\left[
\begin{array}{r}
0 \\
0 \\
1 \\
0
\end{array}
\right] ,  \notag
\end{eqnarray}
\begin{eqnarray}
|11\rangle  &=&|1\rangle \otimes |1\rangle =\left[
\begin{array}{r}
0 \\
1
\end{array}
\right] \otimes \left[
\begin{array}{r}
0 \\
1
\end{array}
\right] =\left[
\begin{array}{r}
0 \\
0 \\
0 \\
1
\end{array}
\right] .  \notag
\end{eqnarray}
Thus, from (\ref{cnot}) and (\ref{vectors}), the matrix representation
$U_{\mbox{\tiny CNOT}}$ of the CNOT gate is
\begin{equation*}
U_{\mbox{\tiny CNOT}}=\left[
\begin{array}{rrrr}
1 & 0 & 0 & 0 \\
0 & 1 & 0 & 0 \\
0 & 0 & 0 & 1 \\
0 & 0 & 1 & 0
\end{array}
\right] .
\end{equation*}

\begin{figure}
    \setcaptionmargin{.5in}
    \centering
    \psfrag{i}{$\sket {i}$}
    \psfrag{s}{$\sket {\s}$}
    \psfrag{xisigma}{$X^i \sket {\s}$}
    \includegraphics{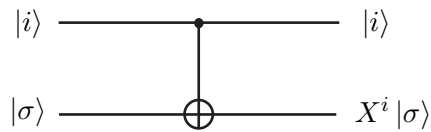}
    \caption{CNOT gate.}
    \label{fig016}
\end{figure}
Fig. \ref{fig016} describes the CNOT gate, where $\sket i$ is
either $\sket 0$ or $\sket 1$. The figure could lead one to think
that the output is always non-entangled, but that is not true,
since if the first qubit is in a more general state given by
$a\sket 0 + b \sket 1$, then the output will be $a\sket 0 \sket
\s+ b \sket 1 X \sket \s$, which is entangled in general.

CNOT and one-qubit gates form a universal set of gates. This means
that any other gate, operating on $2$ or more qubits can be
written as compositions and direct products of CNOT and one-qubit
gates \cite{barenco}.

We have seen two examples of $2$-qubit gates. The general case is
a $4\times 4$ unitary matrix. Gates that are the direct product of
other gates, such as $H\otimes H$, do not produce entanglement. If
the input is non-entangled, the output is not too. On the other
hand, the output of the CNOT gate can be entangled while the input
is non-entangled.

The next gate we consider is the $3$-qubit quantum Toffoli gate. Its action
on the computational basis is given by
\begin{equation*}
|a,b,c\rangle \rightarrow |a,b,c\oplus ab\rangle.
\end{equation*}
The action on a generic state
\begin{equation*}
|\psi \rangle =\sum_{a,b,c=0}^{1}\alpha
_{a,b,c}|a,b,c\rangle =\left[
\begin{array}{c}
\alpha _{000} \\
\vdots  \\
\alpha _{101} \\
\alpha _{110} \\
\alpha _{111}
\end{array}
\right]
\end{equation*}
is obtained by linearity as
\begin{equation*}
|\psi ^{\prime }\rangle =\sum_{a,b,c=0}^{1}\alpha
_{a,b,c}|a,b,c\oplus ab\rangle =\left[
\begin{array}{c}
\alpha _{000} \\
\vdots  \\
\alpha _{101} \\
\alpha _{111} \\
\alpha _{110}
\end{array}
\right] .
\end{equation*}
So, the matrix representation for the Toffoli gate becomes
\begin{equation*}
U_{\mbox{\tiny Toffoli}}=\left[
\begin{array}{rrrrrrrr}
1 & 0 & 0 & 0 & 0 & 0 & 0 & 0 \\ 0 & 1 & 0 & 0 & 0 & 0 & 0 & 0\\ 0
& 0 & 1 & 0 & 0 & 0 & 0 & 0\\ 0 & 0 & 0 & 1 & 0 & 0 & 0 & 0\\ 0 &
0 & 0 & 0 & 1 & 0 & 0 & 0 \\ 0 & 0 & 0 & 0 & 0 & 1 & 0 & 0 \\ 0 &
0 & 0 & 0 & 0 & 0 & 0 & 1\\ 0 & 0 & 0 & 0 & 0 & 0 & 1 & 0
\end{array}
\right] .
\end{equation*}

Further details about quantum circuits can be found in
\cite{barenco,chuang}.

\section{Grover's Algorithm}

Suppose we have an unstructured database with $N$ elements.
Without loss of generality, suppose that the elements are numbers
from $0$ to $N-1$. The elements are not ordered. Classically, we
would test each element at a time, until we hit the one searched
for. This takes an average of $N/2$ attempts and $N$ in the worst
case, therefore the complexity is $O(N)$. As we will see, using
Quantum Mechanics only $O(\sqrt{N})$ trials are needed. For
simplicity, assume that $N=2^n$, for some integer $n$.

Grover's algorithm has two registers: $n$ qubits in the first and
one qubit in the second. The first step is to create a
superposition of all $2^n$ computational basis states $\{
\sket{0}, ..., \sket{2^n-1} \}$ of the first register. This is
achieved in the following way. Initialize the first register in
the state $\sket{0,...,0}$ and apply the operator $H^{\otimes n}$
\bea \sket{\psi} & = & H^{\otimes n} \sket{0,...,0} \non \\
 & = & (H \sket{0})^{\otimes n} \non \\
 & = & \left(\frac{\sket{0} + \sket{1}}{\sqrt{2}}\right)^{\otimes n} \non \\
 & = & \frac{1}{\sqrt{N}} \sum_{i=0}^{N-1} \sket{i} .\label{psi}
\eea
$\sket\psi$ is a superposition of all basis states with equal
amplitudes given by $1/\sqrt{N}$. The second register can begin
with $\sket 1$ and, after a Hadamard gate is applied, it will
be in state $\sket - = (\sket{0} - \sket{1})/\sqrt{2}$.

Now define $f:\{0,...,N-1\}\rightarrow\{0,1\}$ as a function which
recognizes the solution:
\be
\label{f} f(i) = \left\{
\begin{array}{l}
1 \mbox{ if $i$ is the searched element ($i_0$)} \\ 0 \mbox{
otherwise.}
\end{array}
\right.
\end{equation}
This function is used in the classical algorithm. In the quantum
algorithm, let us assume that it is possible to build a linear
unitary operator also dependent on $f$, $U_f$, such that
\be
\label{U} U_f \left( \sket{i}\sket{j} \right) = \sket{i}\sket{j
\oplus f(i)}.
\end{equation}
$U_f$ is called {\em oracle}. In the above equation, $\sket i$
stands for a state of the first register, so $i$ is in $\{0,...,
2^n-1\}$, $\sket j$ is a state of the second register, so $j$ is
in $\{0,1\}$, and the sum is modulo 2. It is easy to check that
\bea U_f \left( \sket{i}\sket{-} \right) & = & \frac{U_f \left(
\sket{i}\sket{0} \right) - U_f \left( \sket{i}\sket{1}\right)
}{\sqrt{2}} \non
\\
 & = & \frac{\sket{i}\sket{f(i)}
- \sket{i}\sket{1 \oplus f(i)}}{\sqrt{2}} \non \\
 & = & (-1)^{f(i)} \sket{i} \sket{-}. \label{ufi-}
\eea
In the last equality, we have used the fact that
\be
1 \oplus f(i) = \left\{
\begin{array}{l}
0 \mbox{  for $i=i_0$} \\ 1 \mbox{  for $i \neq i_0$.}
\end{array}
\right.
\end{equation}

Now look at what happens when we apply $U_f$ to the superposition
state coming from the first step, $\sket\psi \sket -$. The state
of the second register does not change. Let us call $\sket
{\psi_1}$ the resulting state of the first register:
\bea \sket{\psi_1} \sket - & = & U_f \left( \sket \psi \sket -
\right) \non
\\
 & = & \frac{1}{\sqrt{N}} \sum_{i=0}^{N-1}
U_f \left( \sket i \sket - \right) \non \\
 & = & \frac{1}{\sqrt{N}} \sum_{i=0}^{N-1}
(-1)^{f(i)} \sket i \sket - . \label{psi1}
\eea
$\sket{\psi_1}$ is a superposition of all basis elements, but the
amplitude of the searched element is negative while all others are
positive. The searched element has been marked with a minus sign.
This result is obtained using a feature called {\it quantum
parallelism}. At the quantum level, it is possible ``to see'' all
database elements simultaneously. The position of the searched
element is known: it is the value of $i$ of the term with negative
amplitude in (\ref{psi1}). This quantum information is not fully
available at the classical level. A classical information of a
quantum state is obtained by practical measurements, and, at this
point, it does not help if we measure the state of the first
register, because it is much more likely that we obtain a
non-desired element, instead of the searched one. Before we can
perform a measure, the next step should be to increase the
amplitude of the searched element while decreasing the amplitude
of the others. This is quite general: quantum algorithms work by
increasing the amplitude of the states which carry the desired
result. After that, a measurement will hit the solution with high
probability.

\begin{figure}
    \setcaptionmargin{.5in}
    \centering
    \psfrag{0}{$\sket 0$}
    \psfrag{1}{$\sket 1$}
    \psfrag{psiin}{\scriptsize $\sket {\psi_{in}}$}
    \psfrag{psi}{\scriptsize $\sket \psi$}
    \psfrag{psiG}{\scriptsize $\sket {\psi_G}$}
    \psfrag{psig2}{\scriptsize $\sket{\psi_{G^2}}$}
    \includegraphics[width=5.5in]{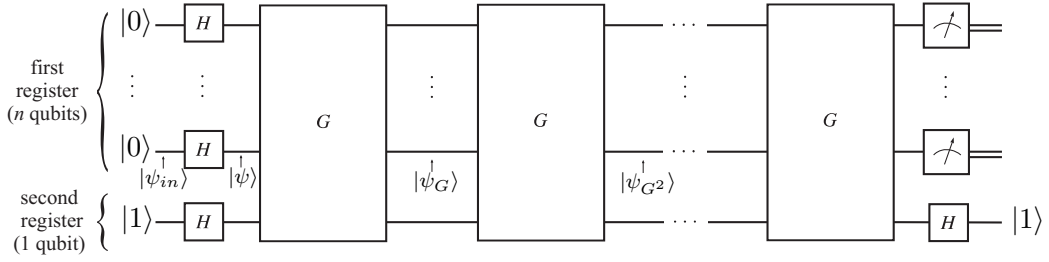}
    \caption[text for list of figures]{Outline of Grover's algorithm.}
    \label{fig08}
\end{figure}

\begin{figure}
    \setcaptionmargin{.5in}
    \centering
    \psfrag{-}{$\sket -$}
    \psfrag{+}[][]{$\frac{\sket 0 + \sket 1}{\sqrt 2}$}
    \psfrag{2psipsi-I}[][]{$2\sket \psi \sbra \psi - I$}
    \psfrag{psi1}[][]{$\sket {\psi_1}$}
    \psfrag{psi}[][]{$\sket \psi$}
    \psfrag{psig}[][]{$\sket {\psi_G}$}
    \psfrag{oracle}[][]{Oracle}
    \includegraphics{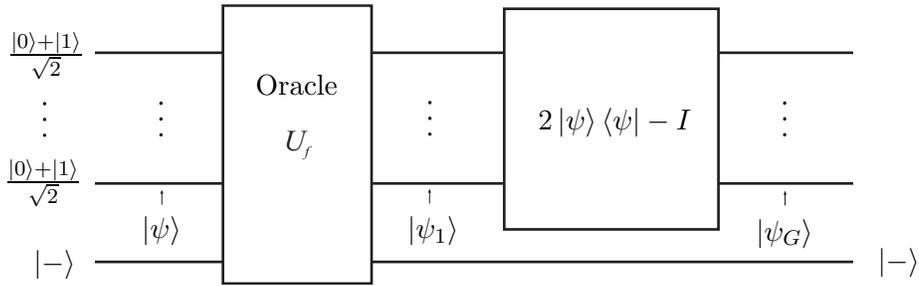}
    \caption{One Grover iteration ($G$). The states of the first register
correspond to the first iteration.}
    \label{fig09}
\end{figure}

Now we shall work out the details by introducing the circuit for
Grover's algorithm (Fig. \ref{fig08}) and analyzing it step by
step.
The unitary operator $G$ is applied $O(\sqrt N )$ times. The exact
number will be obtained later on. The circuit for one Grover
iteration $G$ is given in Fig. \ref{fig09}.
The states $\sket \psi$ and $\sket{\psi_1}$ are given by Eqs.
(\ref{psi}) and (\ref{psi1}), respectively. The operator $2 \sket
\psi \sbra \psi - I $ is called inversion about the mean for
reasons that will be clear in the next section. We will also show
how each Grover operator application raises the amplitude of the
searched element. $\sket{\psi_1}$ can be rewritten as
\be
\label{psi1a}
\sket{\psi_1} = \sket \psi - \frac{2}{\sqrt{2^n}} \sket{i_0},
\end{equation}
where $\sket{i_0}$ is the searched element. $\sket{i_0}$ is
a state of the computational basis. Note that
\be
\label{psii}
\sVEV{\psi | i_0} = \frac{1}{\sqrt{2^n}}.
\end{equation}
Let us calculate $\sket{\psi_G}$ of Fig. \ref{fig08}. Using Eqs.
(\ref{psi1a}) and (\ref{psii}), we obtain
\bea
\label{psig}
\sket{\psi_G} & = & \left(2 \sket \psi \sbra \psi - I \right)
\sket{\psi_1} \non \\
 & = & \frac{2^{n-2} - 1}{2^{n-2}} \sket\psi + \frac{2}{\sqrt{2^n}}
\sket{i_0}.
\eea
This is the state of the first register after one application of
$G$. The second register is in the state $\sket - $.

\section{Geometric Representation}
\label{geo}

All the operators and amplitudes in Grover's algorithm are
real. This means that all states of the quantum computer
live in a real vector subspace of the Hilbert space. This
allows a nice geometrical representation taking $\sket{i_0}$
and $\sket \psi$ as base vectors (non-orthogonal basis).

\begin{figure}
    \setcaptionmargin{.5in}
    \centering
    \psfrag{i0}[][]{$\sket{i_0}$}
    \psfrag{psi}[][]{$\sket{\psi}$}
    \psfrag{psi1}[][]{$\sket{\psi_1} = U_f \sket \psi$}
    \psfrag{psig}[][]{$\sket{\psi_G} = G \sket \psi$}
    \psfrag{q}[][]{$\q$}
    \includegraphics[height=6cm]{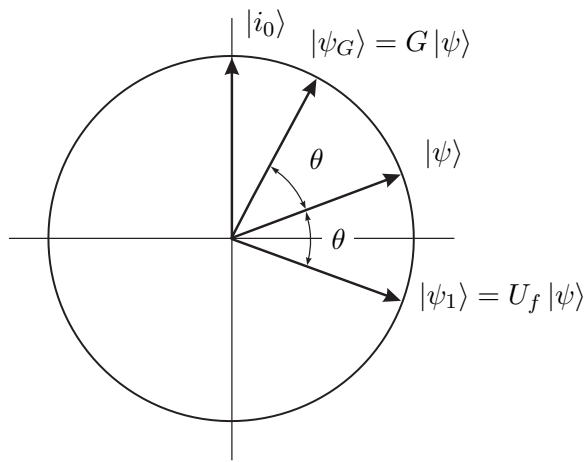}
    \caption{The state of the first register lives in the real
vector space spanned by $\sket {i_0}$ and $\sket \psi$. We take
these states as a basis to describe what happens in Grover's
algorithm.}
    \label{fig010}
\end{figure}

In Fig. \ref{fig010} we can see the vectors $\sket{i_0}$ and
$\sket \psi$. They form an angle smaller than $90^o$ as can be
seen from Eq. (\ref{psii}), since $ 0 < \sVEV{\psi | i_0} < 1 $.
If $n$ is large, then the angle is nearly $90^o$. We can think
that $\sket \psi$ is the initial state of the first register, and
the steps of the computation are the applications of the unitary
operators $U_f$ and $2 \sket\psi \sbra\psi - I$. Then $\sket \psi$
will rotate in the real plane spanned by $\sket\psi$ and
$\sket{i_0}$, keeping the unit norm. This means that the tip of
$\sket\psi$'s vector lies in the unit circle.

>From Eqs. (\ref{psi1a}) and (\ref{psii}) we see that $\sket\psi$
rotates $\theta$ degrees clockwise, where (see $\sket{\psi_1}$ in
Fig. \ref{fig010})
\be
\label{cos}
\cos \theta = 1- \frac{1}{2^{n-1}}.
\end{equation}
>From Eq. (\ref{psig}) we see that the angle between
$\sket{\psi_G}$ and $\sket\psi$ is
\be
\cos \theta^\prime = \sVEV{\psi | \psi_G} = 1- \frac{1}{2^{n-1}}.
\end{equation}
So, $\theta^\prime = \theta$ and $\sket{\psi_1}$ rotates $2
\theta$ degrees counterclockwise (in the direction of
$\sket{i_0}$). This explains the placement of $\sket {\psi_G}$ in
Fig. \ref{fig010}. This is a remarkable result, since the
resulting action of $G = \left(2 \sket \psi \sbra \psi - I \right)
U_f$ rotates $\sket \psi$ towards $\sket {i_0}$ by $\theta$
degrees. This means that the amplitude of $\sket{i_0}$ in
$\sket{\psi_G}$ increased and the amplitudes of $\sket i$, $i\neq
i_0$, decreased with respect to their original values in $\sket
\psi$. A measurement, at this point, will return $\sket{i_0}$ more
likely than before. But that is not enough in general, since
$\theta$ is a small angle if $n \gg 1$ (see Eq. (\ref{cos})). That
is why we need to apply $G$ repeatedly, ending up $\q$ degrees
closer to $\sket {i_0}$ each time, until the state of the first
register be very close to $\sket{i_0}$, so we can measure.

\begin{figure}
    \setcaptionmargin{.5in}
    \centering
    \psfrag{i0}[][]{$\sket{i_0}$}
    \psfrag{psi}[][]{$\sket{\psi}$}
    \psfrag{psi1}[][]{$\sket{\psi_1}$}
    \psfrag{sigma}[][]{$\sket{\sigma}$}
    \psfrag{sigma1}[][]{$\sket{\sigma_1}$}
    \psfrag{gsigma}[][]{$G \sket{\sigma}$}
    \psfrag{q}[][]{$\q$}
    \psfrag{a1}[][]{\small $\a_1$}
    \psfrag{a2}[][]{\small $\a_2$}
    \includegraphics[height=6cm]{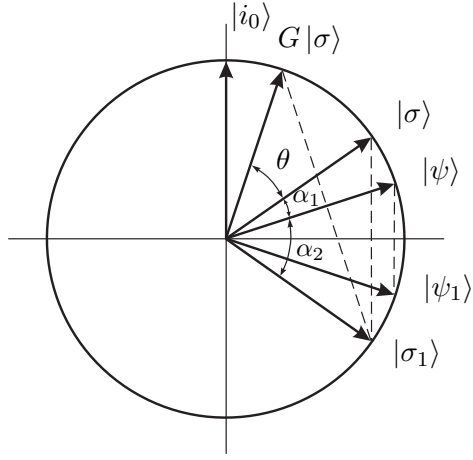}
    \caption{A generic vector $\sket \s$ is reflected around the horizontal
axis by the application of $U_f$, yielding $\sket {\s_1}$. Then, the
reflection of $\sket {\s_1}$ about the mean $\sket \psi$ gives $G\sket\s$,
which is $\q$ degrees closer to $\sket {i_0}$ (vertical axis).}
    \label{fig011}
\end{figure}

Now we show that further applications of $G$ also rotate the state
of the first register by $\theta$ degrees towards $\sket{i_0}$.
The proof is quite general: suppose that $\sket \s$ is a unit
vector making an angle $\a_1$ with $\sket \psi$, as in Fig.
\ref{fig011}.
Let $\sket{\s_1}$ be the state of the first register after the
application of $U_f$ on $\sket \s \sket -$. $U_f$ changes the
sign of the component of $\sket \s$ in the direction of $\sket
{i_0}$. So $\sket{\s_1}$ is the reflection of $\sket \s$ around
the horizontal axis. Let $\a_2$ be the angle between $\sket\psi$
and $\sket{\s_1}$. Let us show that $G \sket \s$ lies in the
subspace spanned by $\sket{i_0}$ and $\sket \psi$:
\bea
\label{Gsigma}
G \sket \s & = & \left(2 \sket \psi \sbra \psi -
I \right) U_f \sket \s \non \\
 & = & 2 \sVEV{\psi | U_f | \s} \sket \psi - \sket{\s_1} \non \\
 & = & 2 \cos \a_2 \sket \psi - \sket{\s_1}.
\eea
We have omitted the state $\sket - $ of the second register
in the above calculation for simplicity. $\sket{\s_1}$ lies
in the subspace spanned by $\sket{i_0}$ and $\sket \psi$,
then $G \sket \psi$ also does.

Now we prove that the angle between $\sket \s$ and $G \sket \s$ is
$\theta$, which is the angle between $\sket \psi$ and
$\sket{\psi_1}$ (see Fig. \ref{fig011}):
\bea
\sVEV{\s | G | \s} & = & 2 \sVEV{\s | \psi} \cos \a_2
- \sVEV{\s | \s_1} \non \\
 & = & \cos \a_1 \cos \a_2 - \cos ( \a_1+\a_2) \non \\
 & = & \cos ( \a_2 - \a_1 ).
\eea
>From Fig. \ref{fig011} we see that $\a_2 - \a_1$ is $\theta$. From
Eq. (\ref{Gsigma}) we see that $G\sket \s$ is a rotation of $\sket
\s$ towards $\sket{i_0}$ by $\theta$ degrees.

The geometrical interpretation of the operator $2\sket \psi \sbra
\psi -I$ is that it reflects any real vector around the axis
defined by the vector $\sket \psi$. For example, in Fig.
\ref{fig011} we see that $G\sket \s = (2\sket \psi \sbra \psi -I)
\sket {\s_1}$ is the reflection of $\sket{\s_1}$ around $\sket
\psi$. $2\sket \psi \sbra \psi -I$ is called inversion about the
mean for the following reason. Let $\sket \s = \sum_{i=0}^{2^n-1}
\s_i \sket i$ be a generic vector and define $\sVEV \s =
\sum_{i=0}^{2^n-1} \s_i$ (mean of the amplitudes of $\sket \s$).
Defining
\be
\sket{\s^\prime} = \sum_{i=0}^{2^n-1} (\s_i -\sVEV \s) \sket i,
\ee
results
\be
(2\sket \psi \sbra \psi -I) \sket {\s^\prime} = - \sket
{\s^\prime}.
\ee
The above equation shows that a vector with amplitudes $\s_i
-\sVEV \s$ is transformed to a vector with amplitudes $-(\s_i
-\sVEV \s)$. Note that $\sket {\s^\prime}$ is not normalized, but
this is irrelevant in the above argument because the amplitudes of
$\sket \s$ and $\sket {\s^\prime}$ only differ by a minus sign.

$U_f$ also has a geometrical interpretation, which can be seen
from the expression
\be
\label{uf} U_f = I - 2\sket{i_0} \sbra{i_0},
\ee
which yields
\be
\label{Ufi}
U_f \sket i = \left\{
\begin{array}{l}
\sket i, \mbox{ if $i \neq i_0$ } \\ -\sket{i_0}, \mbox{  if
$i=i_0$.}
\end{array}
\right. \ee
Therefore, the above representation for $U_f$ is equivalent to Eq.
(\ref{ufi-}) if we do not consider the state of the second
register. The geometrical interpretation is: $U_f$ reflects a
generic vector about the plane orthogonal to $\sket{i_0}$. This is
what Eq. (\ref{Ufi}) shows for vectors of the computational basis.
The interpretation is valid for a generic vector because $U_f$ is
linear. We have not used Eq. (\ref{uf}) to define $U_f$ before,
because we do not know $i_0$ before running the algorithm. On the
other hand, we assumed that it is possible somehow to use function
$f$ given by Eq. (\ref{f}), and to build $U_f$ as given by Eq.
(\ref{U}).

\section{An Example: Grover for \textbf{$N=8$}}

\label{example}

\begin{figure}
    \setcaptionmargin{.5in}
    \centering
    \psfrag{-}{\small $\sket -$}
    \psfrag{1}{$\sket 1$}
    \psfrag{0}{$\sket 0$}
    \psfrag{2psipsi-I}{\small $2\sket \psi \sbra \psi - I$}
    \psfrag{psi1}{\small $\sket {\psi_1}$}
    \psfrag{psi2}{\small $\sket {\psi_2}$}
    \psfrag{psi3}{\small $\sket {\psi_3}$}
    \psfrag{psi0}{\small $\sket {\psi_0}$}
    \psfrag{psi}{\small $\sket \psi$}
    \psfrag{psif}{\small $\sket {\psi_f}$}
    \includegraphics[width=\textwidth]{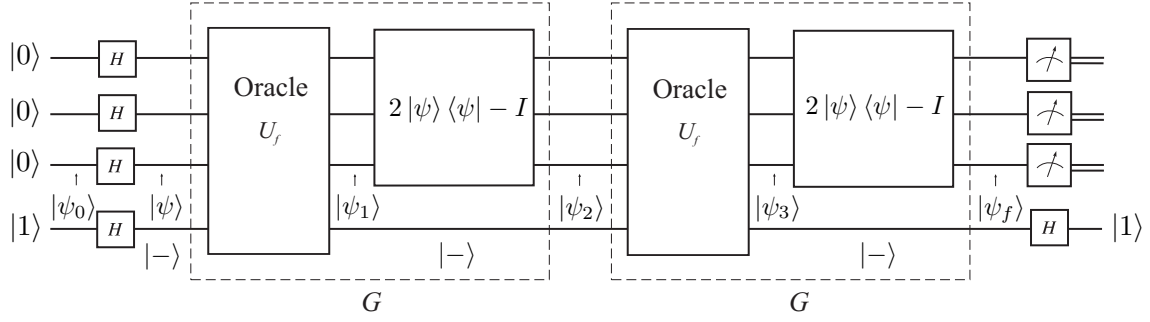}
    \caption{Grover's algorithm for $N=8$.}
    \label{fig012}
\end{figure}

We describe Grover's Algorithm for a search space of 8 elements.
If $N=8$ then $n=3$. There are 3 qubits in the first register and
1 qubit in the second register. For $N=8$, the operator $G$ will
be applied 2 times as we will see in Eq. (\ref{k0ngrande}). The
circuit in this case is given in Fig. \ref{fig012}. The oracle is
queried 2 times. Classically, an average of more than 4 queries
are needed in order to have a probability of success of more than
$1/2$.

\begin{figure}
    \setcaptionmargin{.5in}
    \centering
    \psfrag{u}{$\sket u$}
    \psfrag{101}{$\sket{101}$}
    \psfrag{psi1}{$\sket {\psi_1}$}
    \psfrag{psi2}{$\sket {\psi_2}$}
    \psfrag{psi3}{$\sket {\psi_3}$}
    \psfrag{psi}{$\sket \psi$}
    \psfrag{psif}{$\sket {\psi_f}$}
    \psfrag{q}{$\q$}
    \psfrag{q/2}{\small $\q/2$}
    \includegraphics[height=6cm]{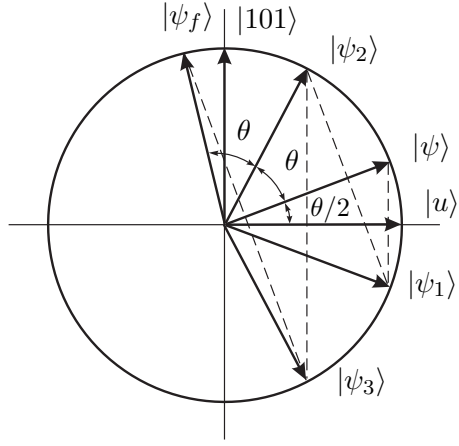}
    \caption{Intermediate states in Grover's algorithm for $N=8$.
Notice how close is $\sket {\psi_f}$ to $\sket{101}$, indicating a
high probability that a measurement will give the searched element. The
value of $\q$ is around $41.4^{\mbox {\tiny o}}$.}
    \label{fig013}
\end{figure}

Let us describe the quantum computer state at each step shown
in the circuit $(\sket{\psi_0},\sket{\psi},$ $\sket{\psi_1},$ $
\sket{\psi_2},$ $\sket{\psi_3},$ and $\sket{\psi_f}$). The initial
state is
\be
\sket{\psi_0} = \sket{000}.
\end{equation}
After applying Hadamard gates,
\be
\sket{\psi} = H^{\otimes 3} \sket{000} = (H \sket{0})^{\otimes 3}
= \frac{1}{2 \sqrt{2}} \sum_{i=0}^{7} \sket i .
\end{equation}
Suppose that we are searching for the element with index 5.
Since $\sket 5 = \sket{101}$,
\bea \label{Uf101} U_f \left( \sket{101} \sket - \right) & = & -
\sket{101} \sket - \non
\\ U_f \left( \sket i \sket - \right) & = & \sket i \sket -   \mbox{ , if } i
\neq 5. \eea
Define $\sket u$ as
\bea \label{u} \sket u & = & \frac{1}{\sqrt{7}}
\sum_{\stackrel{\scriptsize{\mbox{ $i=0$ }}} {\scriptsize{\mbox{
$i \neq 5$ }}}}^7 \sket i \non
\\ & = & \frac {\sket{000}+\sket{001}+\sket{010}+\sket{011}+
\sket{100}+\sket{110}+\sket{111}}{\sqrt{7}}. \eea
Then
\be
\label{psiu}
\sket{\psi} = \frac{\sqrt{7}}{2\sqrt{2}} \sket{u} +
\frac{1}{2 \sqrt{2}} \sket{101}.
\end{equation}
With this result we can see the direction of $\sket{\psi}$ in Fig.
\ref{fig013}.
The value of $\q$ is
\bea
\label{theta}
\q & = & 2 \arccos \left( \frac {\sqrt{7}} {2\sqrt 2} \right) \non \\
& = & \arccos \left( \frac 34 \right) \non \\
& \approx & 41.4^{\mbox {\tiny o}}.
\eea

The next step is
\bea \sket{\psi_1} \sket - & = & U_f \left( \sket \psi \sket -
\right) \non
\\
 & = & \left( \frac {\sket{000}+\sket{001}+\sket{010}+\sket{011}+
\sket{100}-\sket{101}+\sket{110}+\sket{111}}{2\sqrt{2}} \right)
\sket - .
\eea
Note that $\sket {101}$ is the only state with a minus sign.
We can write $\sket{\psi_1}$ as
\be
\label{psi1apsi}
\sket{\psi_1} =  \sket \psi - \frac{1}{\sqrt{2}}\sket{101}
\end{equation}
or
\be
\label{psi1b}
\sket{\psi_1} =  \frac{\sqrt{7}}{2\sqrt{2}} \sket u -
\frac{1}{2 \sqrt{2}}\sket{101}.
\end{equation}
The form of Eq. (\ref{psi1apsi}) is useful in the next step of
calculation, since we have to apply $2 \sket \psi \sbra \psi - I
$. The form of Eq. (\ref{psi1b}) is useful to draw $\sket{\psi_1}$
in Fig. \ref{fig013}. $\sket{\psi_1}$ is the reflection of
$\sket{\psi}$ with respect to $\sket u$.

Next we calculate
\be
\sket{\psi_2} = (2 \sket \psi \sbra \psi - I) \sket{\psi_1}.
\end{equation}
Using Eq. (\ref{psi1apsi}), we get
\be
\sket{\psi_2} = \ha \sket \psi + \frac{1}{\sqrt{2}} \sket{101}
\end{equation}
and, using Eq. (\ref{psiu}),
\be
\sket{\psi_2} =  \frac{\sqrt{7}}{4\sqrt{2}} \sket u +
\frac{5}{4 \sqrt{2}}\sket{101}.
\end{equation}
Let us confirm that the angle between $\sket \psi$ and
$\sket{\psi_2}$ is $\q$:
\be
\cos \q = \sVEV{\psi|\psi_2} = \ha \sVEV{\psi|\psi}
+ \frac{1}{\sqrt{2}} \sVEV{\psi|101} = \frac34,
\end{equation}
which agrees with Eq. (\ref{theta}). This completes one
application of $G$.

The second and last application
of $G$ is similar. $\sket{\psi_3}$ is given by
\be
\sket{\psi_3} =  \frac{\sqrt{7}}{2\sqrt{2}} \sket u -
\frac{5}{4 \sqrt{2}}\sket{101}.
\end{equation}
Using Eq. (\ref{psiu}), we have
\be
\label{psi3}
\sket{\psi_3} = \ha \sket \psi - \frac{3}{2\sqrt{2}} \sket{101}.
\end{equation}
$\sket{\psi_3}$ is the reflection of $\sket{\psi_2}$ with respect
to $\sket{u}$.

The last step is
\be
\sket{\psi_f} = (2 \sket \psi \sbra \psi - I) \sket{\psi_3}.
\end{equation}
Using Eqs. (\ref{psiu}) and (\ref{psi3}), we have
\be
\label{psif}
\sket{\psi_f} =  - \frac{\sqrt{7}}{8\sqrt{2}} \sket u +
\frac{11}{8 \sqrt{2}}\sket{101}.
\end{equation}
It is easy to confirm that $\sket{\psi_f}$ and $\sket{\psi_2}$
form an angle $\q$. Note that the amplitude of the state
$\sket{101}$ is much bigger than the amplitude of any other state
$\sket i$ ($ i \neq 5$) in Eq. (\ref{psif}). This is the way most
quantum algorithms work. They increase the amplitude of the states
that carry the desired information. A measurement of the state
$\sket{\psi_f}$ in the computational basis will project it into
the state $\sket{101}$ with probability
\be
\label{result}
p = \left| \frac{11}{8\sqrt{2}} \right|^2 \approx 0.945.
\end{equation}
The chance of getting the result $\sket{101}$, which reads as
number 5, is around $94,5\%$.

\section{Generalization}

\begin{figure}
    \setcaptionmargin{.5in}
    \centering
    \psfrag{u}{$\sket u$}
    \psfrag{i0}{$\sket {i_0}$}
    \psfrag{gpsi}{$G\sket {\psi}$}
    \psfrag{g2psi}{$G^2\sket {\psi}$}
    \psfrag{g3psi}{$G^3\sket {\psi}$}
    \psfrag{psi}{$\sket \psi$}
    \psfrag{q}{$\q$}
    \psfrag{q/2}{\small $\q/2$}
    \includegraphics[height=6cm]{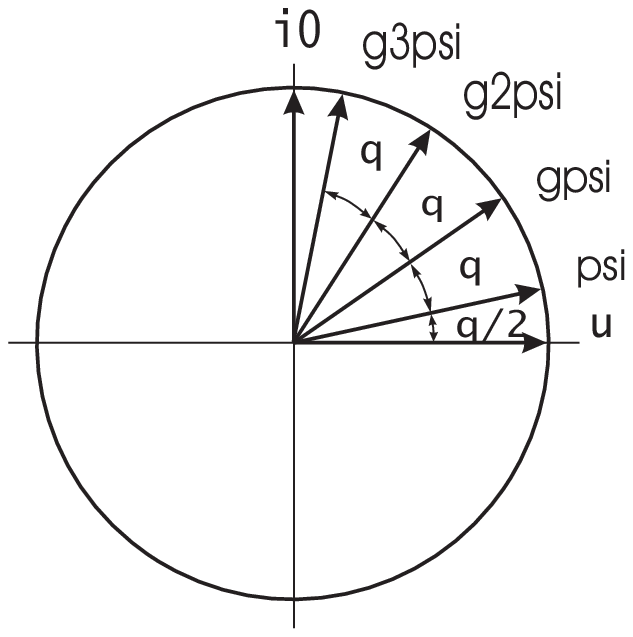}
    \caption{Effect of $G$ on $\sket \psi$.}
    \label{fig014}
\end{figure}

The easiest way to calculate the output of Grover's Algorithm is
to consider only the action of $G$ instead of breaking the
calculation into the action of the oracle ($U_f$) and the
inversion about the mean. To this end, we choose $\sket{i_0}$ and
$\sket u$ as the basis for the subspace where $\sket \psi$ rotates
after successive applications of $G$. $\sket{i_0}$ is the searched
state and $\sket u $ is defined as in Eq. (\ref{u}),
\bea \sket u & = & \frac{1}{\sqrt{N-1}} \sum^{N-1}
_{\stackrel{\scriptsize{\mbox{ $i=0$ }}} {\scriptsize{\mbox{ $i
\neq i_0$ }}}} \sket i \non \\
 & = & \sqrt{\frac{N}{N-1}} \sket \psi -
\frac{1}{\sqrt{N-1}} \sket{i_0} .
\eea
>From the first equation above we easily see that $\sVEV{i_0|u} =
0$, i.e., $\sket{i_0}$ and $\sket{u}$ are orthogonal. From the
second equation we have
\be
\label{psiui0}
\sket \psi = \sqrt{1- \frac{1}{N}} \sket u +
\frac{1}{\sqrt N} \sket{i_0}.
\end{equation}
The state of the quantum computer at each step is
\be
\label{Gkpsi}
G^k \sket \psi = \cos \left(\frac{2k+1}{2} \q \right)
\sket u + \sin \left(\frac{2k+1}{2} \q \right) \sket{i_0},
\end{equation}
where we have dropped the state of the second register since it is
$\sket - $ all the time. Eq.(\ref{Gkpsi}) is obtained from Fig.
\ref{fig014} after analyzing the components of $G^k \sket \psi$.
The value of $\q$ is obtained substituting $k$ for $0$ in Eq.
(\ref{Gkpsi}) and comparing it with Eq. (\ref{psiui0}),
\be
\label{teta}
\q = 2 \arccos \sqrt{1-\frac{1}{N}}.
\end{equation}
Eq.(\ref{Gkpsi}) expresses the fact we proved in section
\ref{geo}, that each application of $G$ rotates the state of the
first register by $\q$ degrees towards $\sket {i_0}$. Fig.
\ref{fig014} shows successive applications of $G$.
\begin{figure}
    \setcaptionmargin{.5in}
    \centering
    \includegraphics[angle=-90,totalheight=2in]{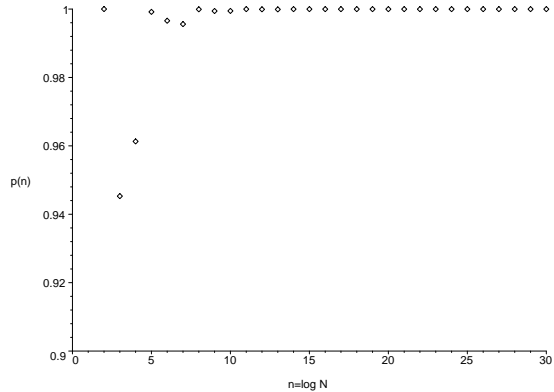}
    \caption{Probability of succeeding as a function of $n$.}
    \label{fig015}
\end{figure}

The number of times $k_0$ that $G$ must be applied
obeys the equation
\be
k_0 \q + \frac{\q}{2} = \frac{\p}{2}.
\end{equation}
Since $k_0$ must be integer, we write
\be
\label{k0}
k_0 = \mbox{round} \left( \frac{\p-\q}{2\q} \right),
\end{equation}
where $\q$ is given by Eq. (\ref{teta}). If $N \gg 1$, by Taylor
expanding Eq. (\ref{teta}), we get $\q \approx 2/\sqrt N$ and from
Eq. (\ref{k0}),
\be
\label{k0ngrande}
k_0 = \mbox{round} \left( \frac \p 4 \sqrt N
\right).
\ee
After applying $G$ $k_0$ times, the probability $p$ of finding the
desired element after a measurement is
\be
\label{p}
p = \sin ^2 \left(\frac{2k_0+1}{2} \q \right).
\end{equation}

Fig. \ref{fig015} shows $p$ for $n$ from 2 to 30. Recall that
$N=2^n$, so for $n=30$ the search space has around 1 billion
elements. For $n=2$ the probability of getting the result is
exactly 1. The reason for this is that Eq. (\ref{teta}) yields $\q
= \p / 3$ and $\sket \psi$ makes an angle $\p / 6$ with $\sket u$.
Applying $G$ one time rotates $\sket \psi$ to $\sket {i_0}$
exactly. For $n=2$, Eq. (\ref{p}) yields $p \approx 0.945$ which
is the result (\ref{result}) of the previous section.

\section{Grover Operator in Terms of the Universal Gates}

\begin{figure}
    \setcaptionmargin{.5in}
    \centering
    \psfrag{psi0}{$\sket {\psi_0}$}
    \psfrag{psi1}{$\sket {\psi_1}$}
    \psfrag{psi2}{$\sket {\psi_2}$}
    \psfrag{psi3}{$\sket {\psi_3}$}
    \psfrag{psi4}{$\sket {\psi_4}$}
    \psfrag{psi5}{$\sket {\psi_5}$}
    \psfrag{iI}[][]{$iI$}
    \includegraphics{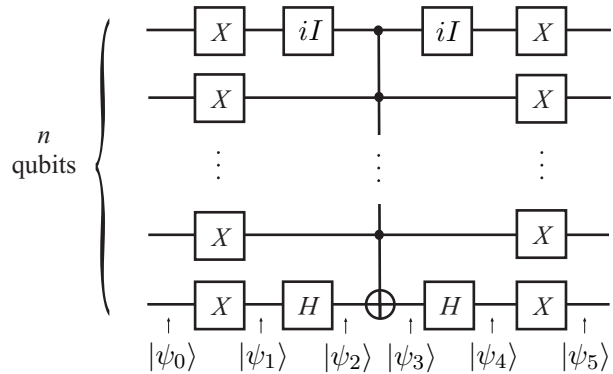}
    \caption{Circuit for $2 \sket 0 \sbra 0 - I$. Note the presence of the
imaginary unit, which does not affect the real character of the
operator.}
    \label{fig017}
\end{figure}

In this section we go in the opposite direction. We decompose $G$
in terms of universal gates, which are CNOT and one-qubit gates.
This decomposition shows how to implement $G$ in practice. Let us
begin by decomposing the inversion about the mean $ 2 \sket \psi
\sbra \psi - I$. Recall that
\be
\sket \psi = H^{\otimes n} \sket 0. \ee
Then
\be
2 \sket \psi \sbra \psi - I = H^{\otimes n} (2 \sket 0 \sbra 0 -
I) H^{\otimes n} . \ee
This equation shows that it is enough to consider the operator $2
\sket 0 \sbra 0 - I$, which inverts a generic vector about the
vector $\sket 0$. The circuit for it is given in Fig.
\ref{fig017}.
One can convince oneself that the circuit gives the correct output
by following what happens to each state of the computational
basis. The input $\sket 0$ is the only one that does not change
signal. The intermediate states as shown in Fig. \ref{fig017} are
\be
\begin{array}{ccrcl}
\sket {\psi_0} & = & \sket 0 \sket 0 & ... & \sket 0 \sket 0 \\
\sket {\psi_1} & = & \sket 1 \sket 1 & ... & \sket 1 \sket 1\\
\sket {\psi_2} & = & i \sket 1 \sket 1 & ... & \sket 1 \sket - \\
\sket {\psi_3} & = & i \sket 1 \sket 1 & ... & \sket 1 (- \sket - )\\
\sket {\psi_4} & = & -i(i \sket 1) \sket 1 & ... & \sket 1 \sket 1 \\
\sket {\psi_5} & = & \sket 0 \sket 0 & ... & \sket 0 \sket 0 .
\end{array}
\end{equation}
The same calculations for the input $\sket j$, $0<j<N$, results in
$- \sket j$ as output.

\begin{figure}
    \setcaptionmargin{.5in}
    \centering
    \psfrag{i1}{$\sket {j_1}$}
    \psfrag{i2}{$\sket {j_2}$}
    \psfrag{in-1}{$\sket {j_{n-1}}$}
    \psfrag{in}{$\sket {j_n}$}
    \psfrag{X}{$X^{j_1 j_2 ... j_{n-1}} \sket {j_n}$}
    \includegraphics{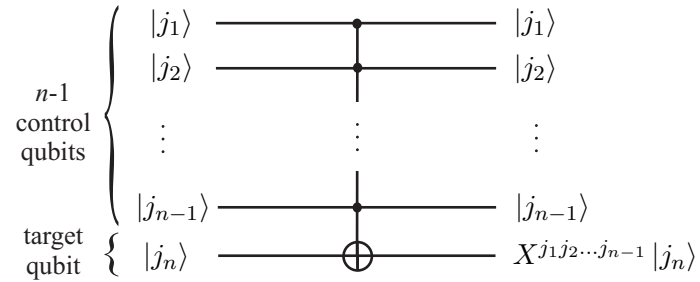}
    \caption{Generalized Toffoli gate.}
    \label{fig018}
\end{figure}
\begin{figure}
    \setcaptionmargin{.5in}
    \centering
    \psfrag{i1}{$\sket {j_1}$}
    \psfrag{i2}{$\sket {j_2}$}
    \psfrag{i3}{$\sket {j_3}$}
    \psfrag{in-1}{$\sket {j_{n-1}}$}
    \psfrag{in}{$\sket {j_n}$}
    \psfrag{X}{$X^{j_1 j_2 ... j_{n-1}} \sket {j_n}$}
    \psfrag{01}{$\sket 0$}
    \psfrag{02}{$\sket 0$}
    \psfrag{0n-3}{$\sket 0$}
    \psfrag{0n-2}{$\sket 0$}
    \includegraphics[width=4.2in]{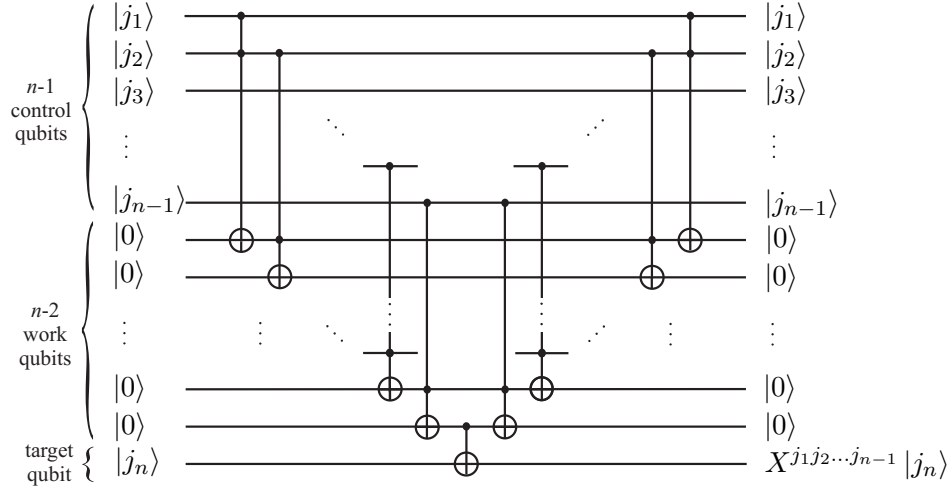}
    \caption{Decomposition of the generalized Toffoli gate
in terms of Toffoli gates.}
    \label{fig019}
\end{figure}
\begin{figure}
    \setcaptionmargin{.5in}
    \centering
    \includegraphics[width=5in]{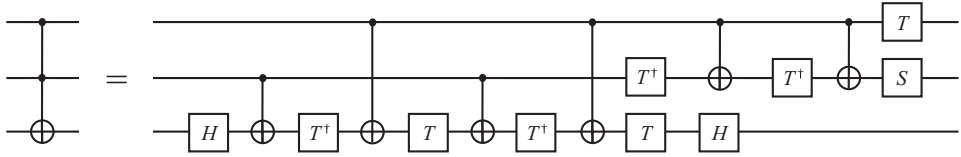}
    \caption{The Toffoli gate in terms of CNOT and one-qubit gates.}
    \label{fig020}
\end{figure}

The only operator in Fig. \ref{fig017} that does not act on single
qubits is the generalized Toffoli gate, which is shown alone in
Fig. \ref{fig018}. The decomposition of the generalized Toffoli
gate in terms of Toffoli gates is given in Fig. \ref{fig019}. The
$n-2$ work qubits are extra qubits whose input and output are
known {\it a priori}. They are introduced in order to simplify the
decomposition. A careful analysis of Fig. \ref{fig019} shows that
the output is the same of the generalized Toffoli gate with the
extra work qubits.

The final step is the decomposition of the Toffoli gate, which is
given in Fig. \ref{fig020}, where $S$ is the phase gate
\be
S = \left[ \begin{array}{cc} 1 & 0 \\
0 & i \end{array} \right]
\end{equation}
and $T$ is the $\p/8$ gate
\be
T = \left[ \begin{array}{cc} 1 & 0 \\
0 & e^{i\p/4} \end{array} \right].
\end{equation}
This decomposition can be verified either by an exhaustive
calculation of tensor products and operator compositions or by an
exhaustive application of operators on basis elements.

\begin{figure}
    \setcaptionmargin{.5in}
    \centering
    \psfrag{m}[][]{$\sket -$}
    \includegraphics[]{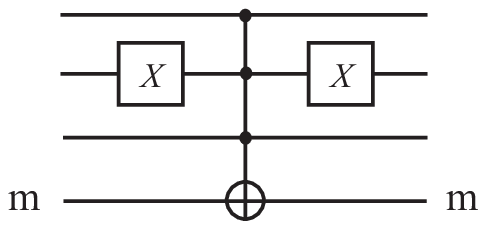}
    \caption{Decomposition of $I - 2 \sket{101}\sbra{101}$, which
simulates $U_f$ that searches number 5.}
    \label{fig021}
\end{figure}

By now one should be asking about the decomposition of $U_f$ in
terms of elementary gates. $U_f$ has a different nature from other
operators in Grover's algorithm, since its implementation depends
on how data is loaded from a quantum memory of a quantum computer.
On the other hand, we have pointed out that $U_f$ can be
represented by $I - 2 \sket{i_0}\sbra{i_0}$ (Eq. (\ref{uf})), if
one knows the answer $i_0$ {\em a priori}. This representation is
useful for simulating Grover's algorithm in a classical computer
to test its efficiency. The operator $I - 2 \sket{i_0}\sbra{i_0}$
is decomposed as a generalized Toffoli gate with $n$ control
qubits, one target qubit in the state $\sket - $, and two
symmetrical $X$ gates in the $i$th qubit, if the $i$th binary
digit of $i_0$ is 0. For example, the operator $U_f$ used in
section \ref{example}, for $N=8$ (see Eq. (\ref{Uf101})) is given
in Fig. \ref{fig021}.

In section 1, we have pointed out that the efficiency of an
algorithm is measured by how the number of elementary gates
increases as a function of the number of qubits. Counting the
number of elementary gates (Figs. \ref{fig08}, \ref{fig09},
\ref{fig017}, \ref{fig019}, and \ref{fig020}), and using Eq.
(\ref{k0ngrande}), we get $\p (17 n - 15) \sqrt{2^n} + n + 2$,
which yields complexity $O(n \sqrt{2^n})$, or equivalently
$\tilde{O}(\sqrt{2^n})$. The notation $\tilde{O}(N)$ means
$O(\mbox{poly}(\log(N))N)$.

\section*{Acknowledgments}

We thank the Group of Quantum Computation at LNCC, in particular,
Drs. F. Haas and G. Giraldi, and the students J.F. Abreu, D.C.
Resende, and F. Marquezino. We thank also Drs. L. Davidovich and
N. Zaguri for stimulating discussions on the subject.


\end{document}